\documentclass[12pt,preprint]{aastex}

\shorttitle{{\it Spitzer} IRS study of Holmberg~II ULX}
\shortauthors{Berghea et al.}

\begin{document}

\title{The First Detection of [O~IV] from an Ultraluminous X-ray Source with {\it Spitzer}. 
I. Observational Results for Holmberg~II ULX}

\author{C. T. Berghea}
\affil{Physics Department, The Catholic University of America, Washington, DC 20064}
\email{79berghea@cua.edu}

\author{R. P. Dudik} 
\affil{United States Naval Observatory, Washington, DC 20392}
\email{rpdudik@usno.navy.mil}

\author{K. A. Weaver and T. R. Kallman}
\affil{Laboratory for High Energy Astrophysics, NASA Goddard Space Flight Center, Greenbelt, MD 20771}

\begin{abstract}
We present the first {\it Spitzer} Infrared Spectrograph (IRS) observations of the [O~IV] 25.89~$\mu$m emission line 
detected from the ultraluminous X-ray source (ULX) in Holmberg~II. This line 
is a well established signature of high ionization, usually associated with AGN. 
Its detection suggests that the ULX has a strong impact on the surrounding gas. 
A {\it Spitzer} high resolution spectral map shows that the [O~IV] is coincident 
with the X-ray position of the ULX.  
The ratios of the [O~IV] to lower ionization lines are similar to those observed in AGN,
suggesting that a strong UV and X-ray source is responsible for the photoionization.
The best {\it XMM-Newton} data is used to model the X-ray band which is then extrapolated into the UV. 
We perform infrared and ultraviolet photometry, and use previously published optical and radio data
to construct the full SED for the ULX and its companion. The preferred model to describe the SED
includes an accretion disk which dominates the soft X-rays but contributes little
at UV and optical wavelengths. The optical counterpart is consistent with a B supergiant as previously suggested in other studies.
The bolometric luminosity of the ULX suggests the presence of an intermediate-mass black hole with mass $>$85~M$_\odot$
for sub-Eddington accretion or, alternatively, a stellar-mass black hole that is accreting at super-Eddington rates.
In a follow-up  second paper we perform detailed photoionization modeling
of the infrared lines in order to constrain the bolometric luminosity of the ULX.

\end{abstract}


\section{INTRODUCTION}

The majority of our knowledge of ultraluminous X-ray sources (ULXs) stems from over 
200 observations of these objects by both {\it Chandra} and {\it XMM-Newton} \citep{liu05}.
The elevated X-ray luminosities of ULXs have raised intriguing questions 
about their nature and the source of the hard X-ray emission.  
A natural explanation for the substantial X-ray luminosities 
is the presence of intermediate-mass black holes \citep[IMBH, e.g.][]{col99}; 
however substantiating evidence for such objects is lacking \citep{wil06, gon06, ber08}.  
Attention has switched to less exotic sources for the emission, such as beaming \citep{king01} 
or super-Eddington accretion from a stellar mass black hole \citep{beg02}, 
both of which explain ULX X-ray properties without an IMBH. 
The idea is not far fetched, since examples of  super-Eddington sources 
have been found in our Galaxy, such as GRS~1915+105 \citep{fen04}, V4641~Sgr \citep{rev02} and possibly SS~433. 
Indeed SS~433 could be an example of both beaming {\it and} super-Eddington accretion, 
the combination of which could easily explain even the most luminous ULXs \citep{beg06, pou07}.

Very little work has been done on ULXs at non X-ray wavelengths, especially in comparison with AGN 
and Galactic stellar-mass black holes. This is in part because, unlike standard AGN, 
the emission from the accretion process itself is harder to detect at these wavelengths.  
Indeed, the X-ray to optical luminosity ratios in ULXs are at least ten times higher 
compared to those seen in AGN \citep{win07}. To make matters worse, there are no ULXs in our own Galaxy 
(though SS~433 is regarded as a ULX by some authors), and the stellar ULX companions 
are very difficult to identify in other galaxies. 

What little multi-wavelength work done on ULXs thus far has been executed primarily 
with high resolution optical spectroscopy. In a few cases, ULXs have been found 
to have an observable effect on their local environment - 
developing ionized bubble nebulae of up to 500~pc diameter \citep{pak02}.
These are much larger than regular supernova remnants and energy estimates show that the bubbles 
are consistent with being blown either by relativistic jets from a ULX, or abnormally powerful supernovae.  
In some cases high-excitation lines have been detected, such as He~II $\lambda$4686 \citep[I.P.~$=$~54.4~eV; e.g.][]{abol07}. 
Only very hot Wolf-Rayet stars can produce this line \citep{sha96}, but the ULX companions are not such stars \citep{abol07}. 
Therefore, the detection of the He~II line must be due to the interaction of the ULX with the surrounding medium. 
Interestingly, SS~433 is also surrounded by a 100~pc radio nebula, W~50 \citep[e.g.][]{fab04}. 

The importance of ionization nebulae for investigating X-ray sources has long been realized, 
beginning with the first detection of nebulae around supersoft sources such as 
LMC X-l \citep{pak86} and CAL~83 \citep{rem95} in the Large Magellanic Cloud (LMC).  
The high-excitation lines observed in nebulae, such as the He~II recombination line, 
can act as giant bolometers for the ionizing UV and X-ray emission, thus allowing 
the intrinsic hard spectral energy distribution (SED) to be inferred. Long-term averaged luminosities 
can be estimated and the UV emission constrained. This provides tremendous value when studying the nature 
of ambiguous ULXs, because the luminosities estimated from ionization 
are independent of X-ray spectral modeling estimates. 
In fact, such luminosities are more accurate, 
because the X-ray-based estimates are in general lower bounds and depend on the viewing angle. 
More importantly, the ionization-estimated luminosities are time-averaged estimates, which is critical 
for variable sources such as ULXs.  This is especially true for ULXs with very few X-ray observations,
since some of these sources can be transient.
On the other hand, photoionization modeling is dependent on many parameters which are often not known very precisely, such as the gas density, geometry, metallicity etc. Contamination from other ionizing sources is also a concern especially in crowded regions.
However, when used together, X-ray analysis and photoionization modeling can provide good constraints on the ULX luminosity. For example, a simple comparison between the two estimates can constrain the beaming factor. 

One of the most interesting examples of an ionized nebula associated with a ULX is located 
in the dwarf galaxy Holmberg~II \citep{pak02,leh05, abol07}. He~II is detected 
in a small nebular region around the ULX, the ``Heel'' of the ``Foot Nebula'' 
\citep[][KWZ hereafter]{kaa04}. An OB star is the likely counterpart (KWZ).  
The morphology of He~II emission and other optical lines is consistent with 
the photoionization of the nebula by the ULX. KWZ confirmed this with CLOUDY simulations. 
\citet{mill05} detected extended radio emission from a $\sim$50~pc diameter region coincident with the position of the ULX. 
The authors conclude that the flux and morphology of the radio emission 
is inconsistent with emission from a SNR or an H~II region, and is probably associated with the ULX activity.

In this paper and a subsequent companion paper we use IR {\it Spitzer} spectroscopic observations for the first time 
to constrain the underlying hard SED of a ULX. The IR observations, 
when combined with previously published optical and X-ray data,  
suggest that the powerful UV and soft X-ray emission from the Holmberg~II ULX photoionizes 
the surrounding medium. 
In Section~2 we present the {\it Spitzer} IRS data analysis and {\it XMM-Newton} X-ray spectral fitting results.
We perform additional photometry measurements with the
Infrared Array Camera (IRAC), the Multiband Imaging Photometer for {\it Spitzer} (MIPS), 
the European Photon Imaging Camera (EPIC) and Optical Monitor (OM) instruments on {\it XMM-Newton}, 
and the Galaxy Evolution Explorer ({\it GALEX}). In Section~3 we analyze the possible contribution
of self-irradiated accretion disk to the UV and optical fluxes.
In Section~4 we use our data and other published data to construct an SED of the Holmberg~II ULX. 
This is also used in the companion paper to further constrain the X-ray spectrum.
In Sections~5 and 6 we discuss the significance of the [O~IV] line detection,
compare the IR line ratios with those of AGN and present the final observational results.
A follow-up paper will present detailed photoionization analysis of the {\it Spitzer} data.

\section{OBSERVATIONS AND DATA ANALYSIS}

We begin this section with a description of the {\it Spitzer} IRS \citep{hou04} observation of the Holmberg~II ULX.
We next present multi-wavelength photometry
and X-ray spectroscopy used in the subsequent section to construct the spectral energy distribution (SED) 
for the ULX.

\subsection{{\it Spitzer} IRS Observations}

The data presented here are archival {\it Spitzer} data, taken on 2004 October 4,
as part of the Legacy Program {\it Spitzer Infrared Nearby Galaxies Survey} \citep[SINGS Program~193, see][]{ken03}. 
The IRS spectral map is described by three parallel pointings and five perpendicular 
at half-slit steps, for both Short High (SH, in the range 9.9-19.6~$\mu$m) 
and Long High (LH, in the range 18.7 $-$ 37.2~$\mu$m) modules.  
The total exposure time was 240~s for both SH and LH observations.   
The data were preprocessed by the IRS pipeline (version 11) at the {\it Spitzer} Science Center 
prior to download. Spectral maps were constructed using ``BCD-level'' processed data 
in conjunction with CUBISM (CUbe Builder for IRS Spectra Maps) 
version 1.5\footnote{Available at: http://ssc.spitzer.caltech.edu /archanaly/contributed/cubism/} 
\citep{smith07, ken03}. The absolute calibration error on fluxes obtained with CUBISM is $\sim$25\% \citep{smith04}.

We adopt a distance of 3.05~Mpc to Holmberg~II (KWZ, 1$\arcsec$ is 15~pc).  
The LH maps obtained with CUBISM have intrinsic sizes of 44.6$\arcsec$ by 28.8$\arcsec$ (670~pc by 430~pc), 
while the SH maps are 27$\arcsec$ by 13.6$\arcsec$ (405~pc by 205~pc). 
However, in order to avoid aperture effects, the spectra presented here 
were extracted from like-sized apertures. Thus the SH and LH spectral data result 
from an equally sized aperture of 23$\arcsec$ by 13.6$\arcsec$ (340~pc by 205~pc), 
which is only slightly smaller than the full SH aperture. 
We also extract spectra from the smallest aperture permitted by CUBISM:  
an 8.9$\arcsec$ by 8.9$\arcsec$ aperture centered on the ULX (i.e. 2~$\times$~2 LH pixels, or 130~pc by 130~pc). 
We will refer to these apertures as the standard aperture and the small aperture, respectively.

\begin{figure}
\epsscale{1}
\plottwo{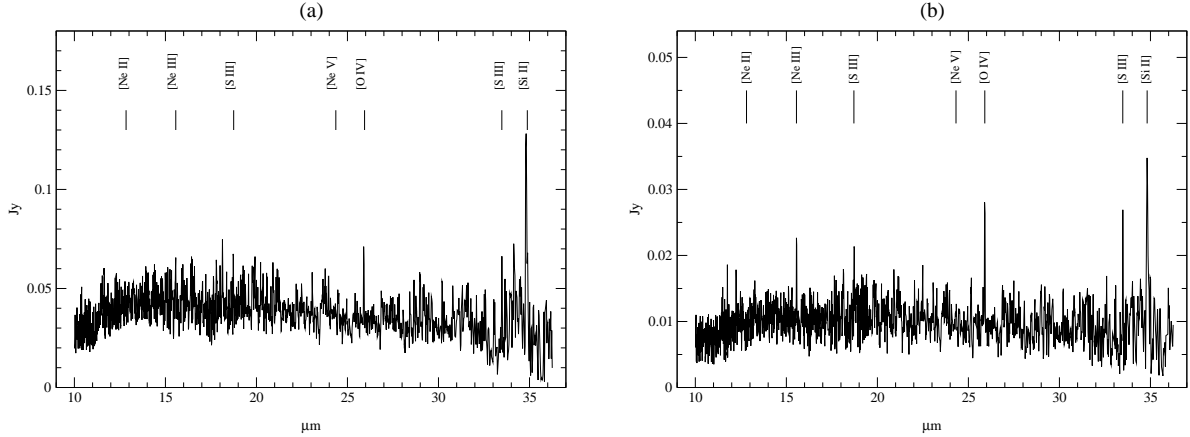}{figure1b.eps}
\caption{
Full IRS spectra obtained from matched apertures with CUBISM from the two apertures shown in Fig.~\ref{maps} :
a) The standard aperture, just slightly shorter than the SH aperture.
b) The small aperture, this is the smallest extraction aperture possible for matched spectra, 4 LH pixels. 
By comparing the spectra it is evident that the [O~IV] is clearly concentrated in the small aperture.
} \label{spectra}
\end{figure}

\begin{figure}
\epsscale{0.8}
\plotone{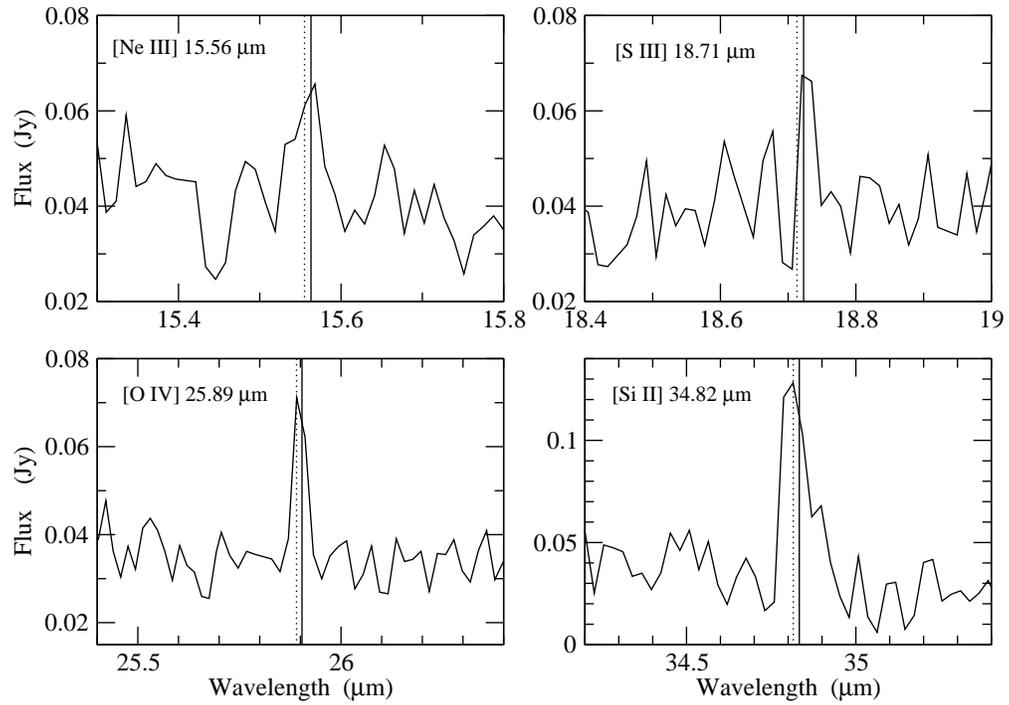}
\caption{
Details of the detected infrared lines in the spectrum of Fig.~\ref{spectra}a. 
The vertical lines show the theoretical center for the lines at rest (dotted) and red-shifted (solid).
We used the recessional velocity adopted by KWZ (157~km~s$^{-1}$).
} \label{zoomed}
\end{figure}

The high-resolution spectra (10~$\mu$m$-$37~$\mu$m) for the two apertures are plotted in Figure~\ref{spectra}.  
Since the spectra are extracted from matched apertures, the spectral lines 
in both the SH and LH modules are properly calibrated and the continuum is properly matched. 
In Figure~\ref{zoomed} we show in detail the 4 detected lines, with the theoretical line centers.
We used SMART \citep[Spectroscopic Modeling Analysis and Reduction Tool, see][]{hig04} version 6.2.6 to measure line fluxes.
Only 4 lines were detected, the fluxes and upper limits for non-detections are presented in Table~\ref{table1}.
Here we show the results for both the standard aperture and the small aperture. 
We note that the [O~IV] line is detected at a signal-to-noise ratio $>$~10.
The spectral lines for the Holmberg~II galaxy are also presented in \citet{dale08}.   
The focus of that paper was not the ULX in Holmberg~II, and as a result, the extraction region 
for the relevant spectral lines is larger than the apertures presented here.  
However, with this in mind, we find that our line fluxes agree reasonably well 
with those presented in \citet{dale08}.

The SH and LH maps for the 4 detected lines are shown in Figure~\ref{maps}.
A false-color Hubble Space Telescope ({\it HST}) image shows the Foot nebula in H$\alpha$ emission,
together with the standard and small extraction apertures. 
Because of the poor spatial resolution of the {\it Spitzer} spectral maps, 
the emission for most lines is difficult to disentangle from the nearby star-forming regions in the Foot nebula. 
This is especially true of the low ionization lines ([Ne~III], [S~III], and [Si~II]).  
The morphologies of these maps indicate that the emission is likely coming from both the ULX 
and the surrounding star forming regions. However, while the low ionization lines [Si~II] and [S~III] are very extended, 
the [O~IV] emission is visibly compact. The peak of the of the [OIV] emission 
is coincident with the {\it Chandra} position for the ULX, with the majority of the emission 
concentrated in the 2$\times$2 pixel region, overlapping with the X-ray point source. 
There is a slight extension to the west and south that is visible on the spectral map 
and which may imply that the [OIV] emission is extended. 
These faint extensions correspond very well with two regions of young star formation 
visible in the {\it HST} image in Figure~\ref{maps}.

KWZ found that the He~II line, which has an ionization potential 
just slightly lower than [OIV] (54.4~eV, compared to 54.9~eV), is also concentrated 
in a small region of 50~pc diameter around the identified optical counterpart of the ULX.
Because the spatial resolution is limited, a detailed comparison of our maps 
with the optical maps presented by KWZ cannot be made. 
However, we find that the morphology of our [O~IV] map appears similar to He~II, 
being brighter in the east and fainter and more extended in the west.

\subsection{X-ray Observations and Spectral Fitting}

We use X-ray spectroscopy to model the X-ray emission from the ULX. 
We also use it to estimate the UV emission by extrapolating the model to lower energies.
X-ray observations of the Holmberg~II galaxy are available from {\it ROSAT}, {\it ASCA}, 
{\it XMM-Newton} and {\it Chandra} \citep{zez99, miy01, kerp02, kaa04, goad06}.  
The X-ray luminosity of the ULX is in excess of 10$^{40}$ erg~s$^{-1}$ in all observations, 
with the exception of a peculiar low and soft state in September 2002, 
when the luminosity dropped to $\sim$5$\times$10$^{39}$ erg~s$^{-1}$ \citep{dew04}. 

To model the X-ray spectrum we used the best available data from the longest ($\sim$100~ks) 
{\it XMM-Newton} EPIC PN observation in the archive (Obs. ID 0200470101, taken on 2003 December 5). 
The data were processed by \citet{win06}. We used XSPEC version 12.4 to model the spectrum between 0.3 and 8.0 keV. 
The X-ray spectrum is the most important component of the total SED since it is the only observed emission 
that directly emanates from the ULX, and is not contaminated by the surrounding star-forming region.  
For this reason we have taken many precautions in modeling the X-ray spectrum.  

The (far) UV emission from the ULX is the least known, but the most important for ionization.
The only available method to estimate the shape of the UV is to extrapolate from X-ray models.
Therefore, it is very important to find the right X-ray model to fit the data, especially in the soft X-ray band.
Three physically different models were used. Each model contains different information about the accretion process.
The first model is very basic and is widely used to model the accretion process for stellar mass black holes, 
ULXs and other accreting objects \citep{mill04, mcc06, win06, fen06}.  
The second is a modification of the first model 
that is more physically realistic when extrapolated to lower energies. 
Finally, the third model is more flexible than the other two models in the UV 
and allows for a tighter fit in the UV while staying within the X-ray error bars.  
The intrinsic (unabsorbed) X-ray luminosity is $>$10$^{40}$ erg~s$^{-1}$ for all models. 
The results from the X-ray spectral fits
are listed in Table~\ref{table2}, and a summary of the models is as follows:

1. {\bf Power-law Plus Multicolor Disk (PLMCD) Model:} 
This is a two-component model with a multi-color accretion disk component \citep[MCD,][]{mit84} 
and a power-law for the hard tail. Following KWZ, the absorption column density was fixed 
at the Galactic value of 3.42$\times$10$^{20}$ cm$^{-2}$, and an additional local absorption 
(N$_H$) model was added. We assume a local metallicity of 0.1~Z$_\odot$ \citep[see][]{kaa04, pil04}. 
This is a very simplistic model commonly used to fit X-ray data for ULXs.  
However, the model falls apart when extrapolated to lower energies, where the power-law dominates.  
The extrapolation is actually physically unrealistic, since it is impossible for the disk component
to Comptonize lower energy photons.
\citet{done06} pointed out that the power-law component in such cases needs to have a break at low energies,
in agreement with the theoretical model where the photons from the disk are up-scattered in the hot corona.
Moreover, if the power-law is produced by another mechanism (e.g. a jet), 
it would not extend down to the soft UV and optical spectrum because 
the fluxes here would be over predicted (see below, Section~4).
Nevertheless, the model continues to be used extensively in X-ray spectral fitting of ULXs.
In conclusion this model does fit the X-ray data well ($\Delta\chi^2$/dof~$=$~1.14), 
but it does not make physical sense when extrapolated to lower energies.
It is shown here as a base, from which alterations can be made,
resulting in  more physically meaningful models.
The following model is one such modification to the PLMCD model, 
and is more physically realistic when extrapolated into the UV.  

2. {\bf Modified PLMCD Model:}   
In order to isolate the intrinsic disk emission from the Comptonized disk emission, 
we fitted the MCD and power-law separately for two different regions of the X-ray spectrum. 
We first fitted the hard tail with a power-law. The spectrum departs from this power-law 
around $\sim$1~keV. Then we fitted an MCD model to the 0.3$-$1.0~keV band.  
The disk temperature (0.38~keV) is slightly higher than other published fits (KWZ obtained $\sim$0.2~keV), 
because our fitting methods are slightly different. Lower temperature accretion disks 
are generally more effective ionizing sources. 
Therefore this higher temperature model may be viewed as a lower bound to the SED in the FUV and soft-Xray bands.
We note that the Modified PLMCD differs from the PLMCD model in that the UV band is predicted 
from the disk component rather than the extrapolated PL component.  
This difference yields a much more physically realistic interpretation for the emission.

3. {\bf Broken Power-Law (BPL) Model:}  
Finally we also fitted the X-ray spectrum to a simple broken power-law model. 
\citet{kaa03} fit a similar model to the ULX in NGC~5408.  
Such a model could be explained if the emission from the companion star were Comptonized by a jet \citep{geo02}.
This model is much more flexible in the UV than either the PLMCD or the Modified PLMCD models, 
and can predict the {\it slope} of the SED in the UV while simultaneously fitting the hard power-law component 
in the X-rays within the error bars of the X-ray measurements. Surprisingly, this simple model 
provides a very good fit to the X-ray data, better than the PLMCD model.
The break energy is at 1.0~keV, the same energy at which we split
the two components in the Modified PLMCD model (2 above).  
Finally, we note that, even if this model has a power-law less steep than the PLMCD model at lower energies,
it is still overpredicting the soft UV and optical emission according to our photometry results (Section~4, below).
A similar overprediction at lower energies was found for this model by \citet{done06} for 
the Galactic binary XTE J1550-564 when at high accretion rates (in the very high state).

\subsection{SED of the ULX: Multi-wavelength Photometry}

\begin{figure}
\epsscale{1.0}
\plottwo{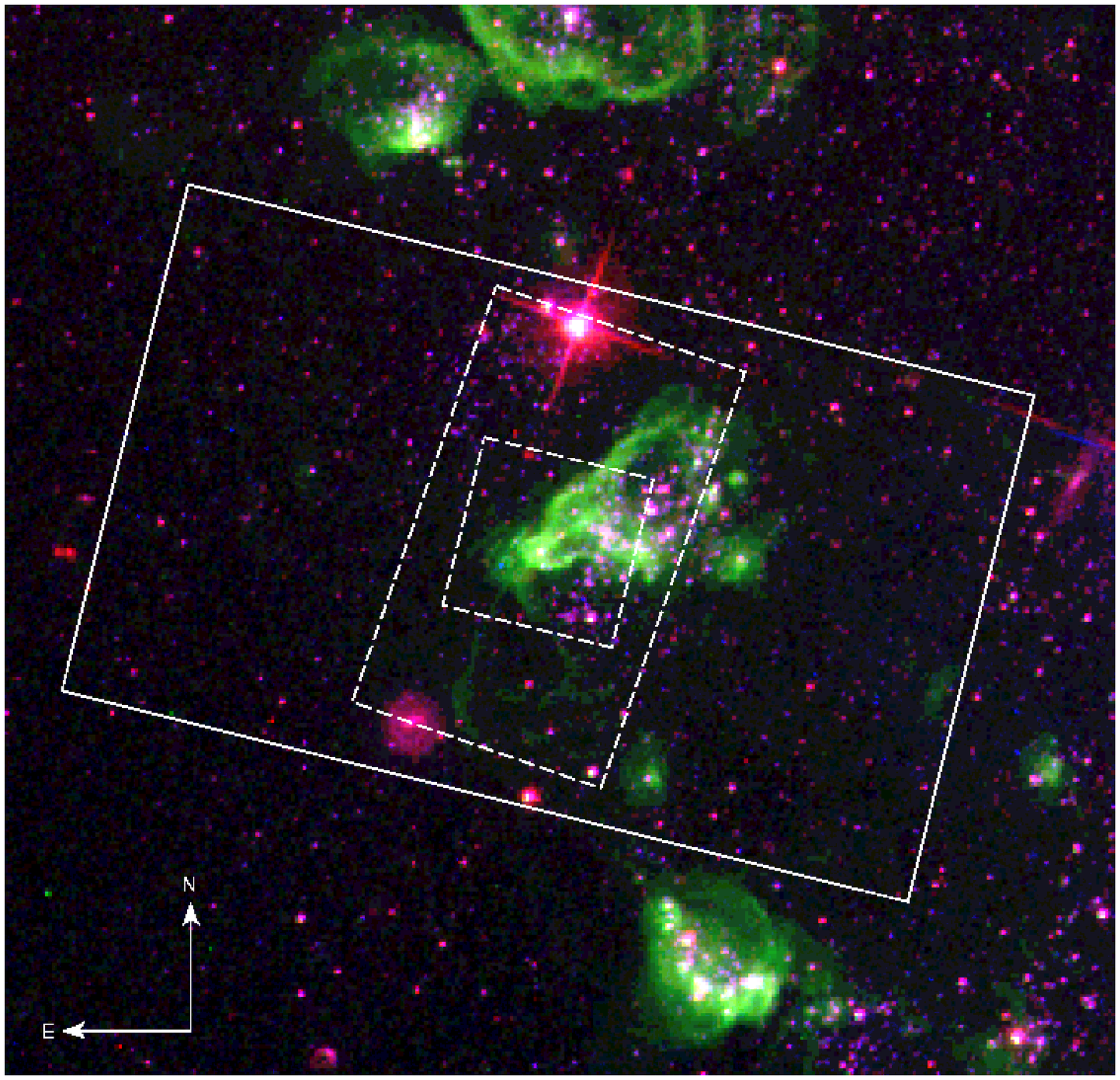}{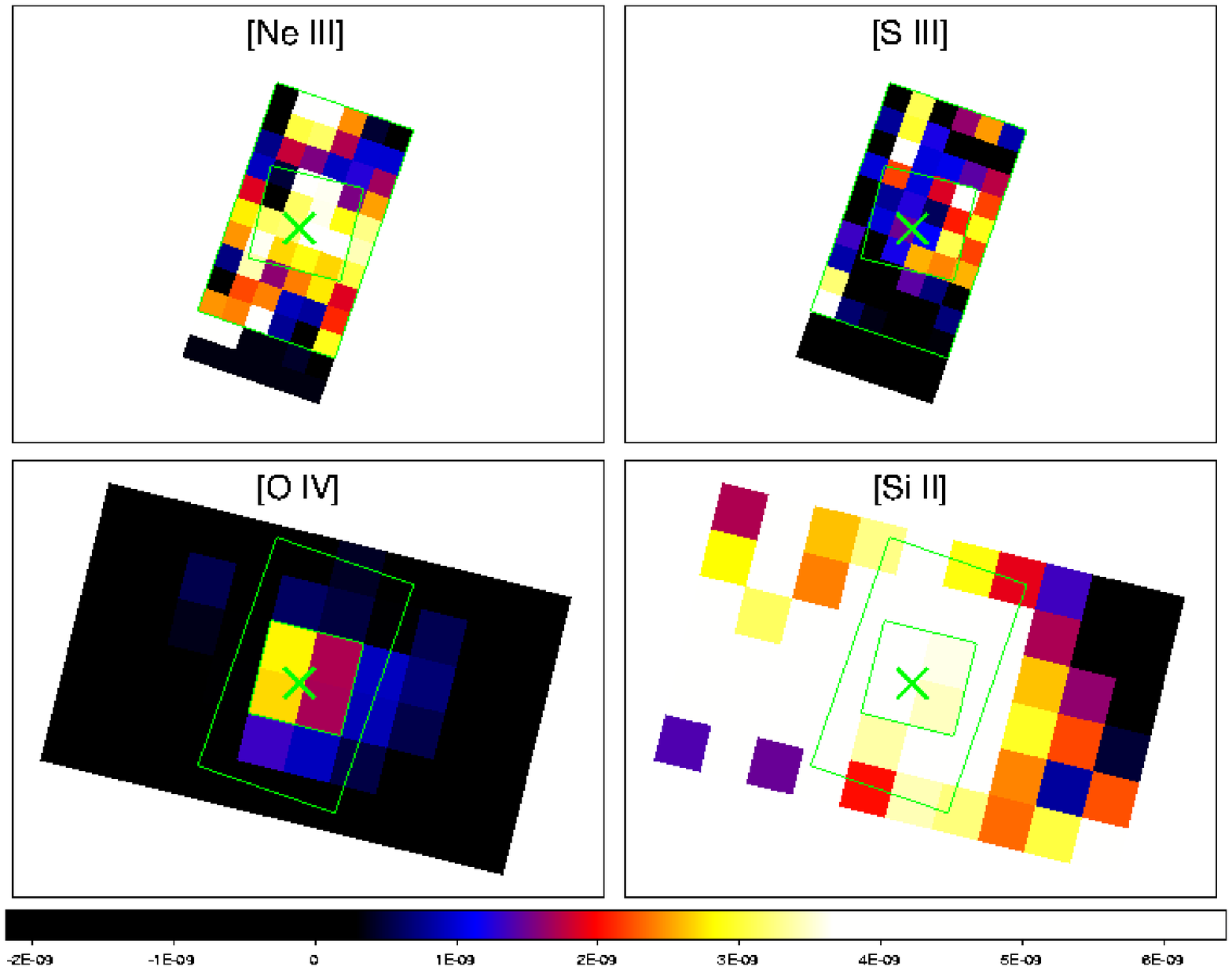}
\caption{
{\it HST} ACS false-color image of the Foot nebula in Holmberg~II (left), 
and the {\it Spitzer} spectral maps of the infrared detected lines (right).
The colors in the {\it HST} image correspond to the following filters: 814W (red), 658N (green, H$\alpha$) and 550M (blue).
The spectral maps are: [Ne III] 15.56~$\mu$m, [S III] 18.71~$\mu$m, [O IV] 25.89~$\mu$m, 
[S III] 33.48~$\mu$m and [Si II] 34.82~$\mu$m. The units on the colorbar are W~m$^{-2}$~sr$^{-1}$.
The apertures used to extract IRS spectra (standard and the small aperture, see Section~2.1), 
are overplotted on all images. In the {\it HST} image we also show 
the whole LH map for comparison with the LH maps shown at right.
The ULX optical counterpart identified by KWZ has the appearance of a bright star in the {\it HST} image,
close to the center of the small aperture, and its position is shown with a green X on the spectral maps. 
It is located at the center of the Heel of the nebula - the round-shaped strong H$\alpha$ emission,
at the center of the small aperture.
An OB association of stars is seen west from the ULX (see also the OM image in Fig.~4).
They  appear to have blown a large bubble in the gas, creating the "Sole" and "Toes" of the Foot nebula.
The low-ionization lines are probably contaminated by these stars.
Some contamination is also produced by the bright foreground star,
seen on the {\it HST} image just inside the standard aperture, close to its northern edge.
} \label{maps}
\end{figure}

In order to constrain ULX multi-wavelength SED, we performed photometry in the IR and UV 
using {\it Spitzer}, {\it XMM-Newton} and {\it GALEX} data.
We also incorporated previously published optical and radio data from {\it HST} and the {\it Very Large Array} ({\it VLA}). 
A short description of the data and the measurements are shown in Table~\ref{table3}.
All our photometric measurements were extracted from an aperture of 50~pc diameter (3.3$\arcsec$).
We used the optical counterpart identified by KWZ to center the extraction aperture (see Fig.~\ref{maps}).
Aspect correction for the IR and UV images was performed using bright sources in field,
such as the foreground star seen in Figure~\ref{maps}.
The measurements presented in Table~\ref{table3} represent the integrated fluxes over this region, 
including the nebular emission and the contamination from other nearby sources.  
They should therefore be considered upper bounds to the intrinsic ULX contribution.

{\bf IR Photometry:} IRAC and MIPS images (see Fig.~4) were taken as a part of SINGS Program~195. 
We extracted 3.6~$\mu$m and 8.0~$\mu$m fluxes using 50~pc diameter apertures centered on the ULX position.
For the MIPS 24~$\mu$m photometry, we used a 100~pc extraction aperture and then divided the flux by four
to obtain an approximate flux for an equivalent 50~pc aperture. 
We followed the standard extraction procedure and conversion factors 
provided in the {\it IRAC Data Handbook} \footnote{http://ssc.spitzer.caltech.edu/irac/dh/}. 
The pipeline mosaics for IRAC and MIPS are accurate to within 20\%.

{\bf UV Photometry:}  {\it XMM-Newton} Optical Monitor (OM) images of Holmberg~II (Fig.~4) are available in the UVW1 band, 
which is centered at 2910\AA (FWHM 620\AA). We use the mosaicked image created for the Optical Monitor Catalog \citep[OMCat,][]{kuntz08}, 
which has an improved coordinate correction (better than 0.5$\arcsec$) over the automated pipeline. 
To measure the UVW1 count rate we used the {\it XIMAGE} software included in the {\it HEAsoft} package. 
We converted the count rate into flux using the conversion factors listed 
in the {\it XMM-Newton Science Analysis System User's Guide}\footnote{http://xmm2.esac.esa.int/external/xmm\_user\_support/documentation/}.   
We obtained a UVW1 magnitude of 18.42~$\pm$~0.06, which corresponds to a luminosity of 8.3~$\times$10$^{37}$ erg~s$^{-1}$ 
and a flux density of 33.1~$\pm$~0.1~$\mu$Jy. The OM image in Figure~4 shows that the Foot Nebula 
has a strong UV emission which peaks on the OB association of stars seen in the {\it HST} image of Figure~\ref{maps} west of the ULX. 
The total UVW1 flux for this UV bright region (including the ULX and its optical counterpart) is equivalent to 29 O5V stars. 
Figure~\ref{maps} shows that these bright stars have apparently blown a cavity in the NW part of the nebula,
thus producing the ``Sole'' and the ``Toes'' of the Foot nebula. 
We note that these cavities are correlated with the faint extended [O~IV] emission 
seen in the CUBISM spectral maps (Fig.~\ref{maps}). The strongest emsission of H$\alpha$
is however in the Heel that hosts the ULX.

We also used archived {\it GALEX} observations, taken as a part of the {\it GALEX} Nearby Galaxies Survey \citep[NGS,][]{bia06}.  
Photometric fluxes were extracted from the FUV (1350$-$2013~\AA) and NUV (1750$-$2013~\AA) bands with the {\it XIMAGE} tool. 
The {\it GALEX} images have a lower resolution compared to OM, but they show similar features.

{\bf Published Radio Data:}  Finally, we searched the literature for published photometric data for the Holmberg~II ULX. 
Using radio {\it VLA} data, \citet{mill05} measured 1.174~$\pm$~0.0085~mJy at 1.4~GHz and 0.677~$\pm$~0.207~mJy at 4.86~GHz. 
\citet{mill05} found extended emission covering the {\it Chandra} position of the ULX:
3.7$\arcsec$ by 2.7$\arcsec$ at 1.4~GHz and 3.4$\arcsec$ by 1.9$\arcsec$ at 4.86~GHz.
These are comparable to the 50~pc extraction region used in the IR and UV photometry.

{\bf Published Optical Photometry:} In addition, KWZ measured a V-band magnitude 
of 22.04~$\pm$~0.08 for the stellar optical counterpart from observations 
with the Advanced Camera for Surveys (ACS, F550M filter, see Fig.~\ref{maps}) on the {\it HST}. 
The optical flux is likely less contaminated by the nebular emission or emission from other sources
compared to our measurements, because of the much smaller aperture.

{\bf Extinction Corrections:} The Milky Way reddening in the direction of Holmberg~II is E(B$-$V)~$=$~0.032
(from the NASA Extragalactic Database).
KWZ estimated an additional local reddening of E(B$-$V)~$=$~0.07~$\pm$~0.01 based on the X-ray absorption.
Thus the total extinction for R$_V=$~3.1 is A$_V=$~0.32. We show the projected extinction and the extinction-corrected
values for the UV and optical data in the last two columns of table Table~\ref{table3}.
These were calculated using the \citet{car89} extinction curves.
For the infrared data, we estimate from \citet{draine} that the extinction is A$_{\lambda}<$~0.02, and therefore negligible.
We note that the photometric fluxes presented here are likely contaminated by the surrounding nebular emission 
as is evidenced by the {\it HST} images in Figure~\ref{maps}.
Moreover, KWZ caution that the local extinction could in fact occur very close 
to the compact object and therefore not affect our measurements at all.
We will therefore take the (extinction-corrected) photometric measurements as upper bounds.

\section{SELF-IRRADIATION OF THE ACCRETION DISK}

While the MCD accretion disk model is a good approximation of the UV continuum to first order, 
it does not account for self-irradiation of X-ray photons caused by thickening of the accretion disk 
at the outer edges and relativistic effects on the X-rays emitted from the inner regions of the disk.  
In Galactic X-ray binaries, this reflected radiation has been shown to increase the optical luminosity of the accretion disk
by as much as 18\% \citep{san93,vrt90,rev02}. Thus if irradiation were contaminating the UV emission in the Holmberg II ULX, 
such a model must be included in the final fit.  

We therefore checked to see if X-ray irradiation would affect the ionizing UV emission from the Holmberg~II ULX, 
or contribute significantly to the flux of the optical counterpart. Following \citet{san93}, 
we modeled the expected SED of an irradiated disk for black holes of mass 10, 100 and 1000~M$_{\odot}$ 
and with varying accretion disk radii. Here the accretion disk radius is constrained by the size of the Roche lobe, 
and is therefore dependent on the spectral type of the companion star. We find that for stellar mass black holes, 
irradiation affects the accretion disk model significantly for most companion spectral types. 
However, for intermediate black hole masses, irradiation is only important to systems with late-type supergiant companions. 

In Figure~5 two examples of irradiated spectra are plotted.  These represent typical spectra for stellar-mass BHs (10~M$_{\odot}$) 
and IMBHs (1000~M$_{\odot}$). In the latter case, an accretion disk typical for a system with an M-supergiant was used 
to illustrate the effect of irradiation. Only if the IMBH has such companions is the self-irradiation effect significant. 
However, we note that an M-supergiant companion is not consistent 
with the observations for the ULX in Holmberg II (see Section~4 below).  As can be seen from this plot, 
in both cases the irradiation dominates the optical and near-IR spectrum much more than the UV band.  
Thus we conclude that while irradiation may be important to model in the optical spectrum of a system with a stellar mass black hole, 
it is likely not dominant for an IMBH system. Finally we conclude that the irradiation does not affect 
the detected IR lines. Moreover, because the measured fluxes are much higher than the predictions from the accretion disk, self-irradiation is not likely to make a significant contribution to our measured fluxes at any wavelength.

\section{SPECTRAL ENERGY DISTRIBUTION (SED)}

Using the measurements and the X-ray models in Section~2, we construct the SED of the ULX using the IR, optical, 
UV and X-ray photometric measurements. The final SED is plotted with the continuous black line in Figure~6. 
All photometric measurements described in the previous section were extracted from an aperture of $\sim$50~pc 
centered on the ULX (Fig.~4). 
The photometric measurements presented here are likely contaminated 
by the surrounding star-forming regions west of the ULX (see the {\it HST} image in Fig.~\ref{maps}  and the OM image of Fig.~4).
As a result, in constructing the SED we assume these are upper bounds to the ULX emission.

{\bf X-ray to UV Band:}  The UV band is responsible for producing the IR emission lines,
however it is also unfortunately the most ambiguous to fit. 
We found that both X-ray models that have power-law components extending to lower energies
(PLMCD and BPL), over-predict the measured fluxes if extrapolated into UV and optical.
Therefore we use the Modified PLMCD model to extrapolate the X-ray spectrum to lower energies. 
Here, the disk emission can be approximated as a simple power-law with photon index 2/3 
\citep[or energy index -1/3, see][]{mit84}, since this extrapolated disk component  is a superposition 
of numerous black bodies representing the integrated emission from the disk 
at the various temperatures and radii. At optical energies, this model 
predicts a drop in flux due to the finite size of the accretion disk (see Fig.~5).

{\bf Optical Band:}  The disk model was further extrapolated into the optical, 
where the companion star dominates the SED. 
KWZ found that stars within the spectral type range B3Ib to O4V were consistent with optical data.
Based also on the UV fluxes presented here we chose a B2Ib supergiant as the optical counterpart.
The UV fluxes seem to be too low for an O5V star.
The model for the B2Ib star is part of the ATLAS9 Model Atmospheres \citep{cas04}, with a temperature of 18500~K 
and a luminosity of 5$\times$10$^{38}$ erg~s$^{-1}$. Figure~6 shows that the supergiant 
emits very few photons that can impact the ionization. On the other hand, 
an O5V star provides a significant flux above 13.6~eV, but is equally deficient above the He~II edge. 

{\bf Infrared and Radio Band:}  At longer wavelengths, from IR to radio, 
we used a broken power-law to approximate the SED. These photometric data are viewed as upper bounds
since the ULX flux is likely contaminated by the surrounding star-forming region.

\section{DISCUSSION}

\subsection{The SED of the ULX in Holmberg II}

In Section~4 and Figure~6 we constructed the full SED for the Holmberg~II ULX 
using multi-wavelength data and previously published data. The bolometric luminosity for this SED 
is 1.34$\times$10$^{40}$ erg~s$^{-1}$, when the far-UV spectrum is extrapolated 
from the X-rays, using the Modified PLMCD model fit. Table~2 illustrates the degeneracy 
resulting from the three methods for X-ray spectral fitting. All three show a fairly good fit to the X-ray data, 
making it very difficult to distinguish between the different physical interpretations that each model represents.
 
The optical and UV fluxes plotted in Figure~6 can, in principle, come from three different sources: 
the stellar companion, the accretion disk, and from a jet. However, the overall SED shape 
and in particular the fact that the optical flux is much lower than the UV fluxes, 
disfavor the jet scenario as discussed below. On the other hand, the emission from the disk in the Modified PLMCD model 
is too faint at optical and UV wavelengths if extrapolated from X-rays and requires an additional source.  
Previously published optical data and our UV measurements suggest that a $\sim$B2Ib star is the source of these fluxes 
as the companion to the accreting BH. We explore this further in Paper~2.

The fact that the PLMCD model overpredicts the UV and optical fluxes if extrapolated to these wavelengths is expected, 
because the power-law cannot extend too much below the peak of the accretion disk \citep[e.g.][]{done06}. 
On the other hand, it is intriguing that the simple broken power-law model (BPL) provides such a good fit 
to the high-quality X-ray data. Moreover, the location of the break ($\sim$1~keV) in this case 
is in agreement with the jet model. \citet{kaa03} fit a very similar model to another famous ULX in NGC~5408.  
It has been shown that jets can contribute significantly to the X-ray emission, 
both by optically thin synchrotron emission and Compton scattering of the photons from the disk (or the companion) 
by relativistic electrons in the jet \citep{geo02}. However, even if the power-law is less steep 
compared to the PLMCD model, it is still overpredicting the UV and optical fluxes and is therefore not the preferred scenario.

Much better constraints are provided by photoionization modeling of the {\it Spitzer} spectra.  
The [O~IV] line is only produced by high energy photons ($\gtrsim$~54~eV), 
and is clearly associated with the ULX (Fig.~\ref{maps}).  
For instance, the ``classical'' PLMCD model predicts a power-law component with photon index of 2.4 dominating into the UV. 
This model has the strongest emission of the three models in the far UV and soft X-rays, 
and should produce the strongest [O~IV] emission. In this case, photoionization modeling can be used 
to predict the [O~IV] flux predicted from such a model, and this predicted flux can be compared with our observed values. 
In a companion paper (Paper~2) we present the results of such a photoionization analysis in an attempt to distinguish 
between the three X-ray models presented here.

\subsection{Characterization of the Optical/Mid-IR Line Ratios}

The ionization potential of [O~IV] is just above the He~II edge so both likely result from similar regions in the nebula 
and a quantitative comparison between previously published optical observations and the mid-IR observations can be made.  
KWZ found that the He~II recombination line is consistent with photoionization by the ULX.  
The flux of our [O~IV] line is almost three times higher than the He~II flux measured by KWZ (2.7~$\times$~10$^{36}$ erg~s$^{-1}$).

We compared the mid-IR line ratios of the ULX with those in AGN. NGC~4395 is one of the few known galaxies 
that is likely powered by an IMBH \citep{fil03}. It has a bolometric luminosity very close to our ULX, 
and high-excitation emission lines were detected. The narrow-line He~II measured by \citet{kra99} for NGC~4395 
relative to H$\beta$ is 0.12, a value typical for Seyfert1 galaxies \citep{die05}. For the Holmberg~II ULX, 
He~II/H$\beta\approx$~0.2 (KWZ); slightly higher than for NGC~4395. The [O~IV] measured from NGC~4395 
is almost 16 times larger than the He~II flux \citep{kra99}. However, the gas density 
is close to the critical density for this line (10$^4$~cm$^{-3}$) $-$ much higher than the density of the gas 
surrounding the Holmberg~II ULX (KWZ estimate 10~cm$^{-3}$, based on the line surface brightness predicted with CLOUDY). 
The [O~IV] is expected to increase 
as it approaches the critical density, while the He~II is not sensitive to the gas density.  
An even more important difference is the abundance of O, 
which is $\sim$~5 times higher for NGC~4395 \citep{kra99}. The dependence of the [O~IV] on the abundance
is expected to be roughly linear. In spite of these two differences, 
we consider the [O~IV] (and He~II) detected from the Holmberg~II ULX 
to be consistent with photoionization by a high energy source similar to the AGN in NGC~4395.

The infrared [O~IV] line (together with [Ne~V]) is a signature of high ionization 
usually associated with actively accreting black holes \citep[e.g.][]{lutz98}. 
This line is often used to disentangle the starburst emission in composite type galaxies, 
and is a good indicator of the AGN power \citep{gen98,sturm02,sat04,smith04,sat07, mel08}. 
\citet{gen98} and \citet{sturm02} have used the ratio of [O~IV]/[Ne~II] to distinguish between
starbursts and AGN dominated galaxies in the mid-IR. In Figure~\ref{ratios}a we compare
the Holmberg~II ULX with data published by \citet{gen98}. The comparison clearly indicate high ionization 
for the ULX, similar to AGN. 

Other infrared line diagnostics have been proposed by \citet{dale06, dale08} 
to disentangle AGN type ionizing sources from star formation, using the ratios [S~III] 33.48~$\mu$m / [Si~II] 34.82~$\mu$m
and [Ne~III] / [Ne~II]. Figure~\ref{ratios}b shows the four regions defined by \citet{dale06}
in their Figure~5, and their published data for different sources in the SINGS sample.
The Holmberg~II ULX is located well within region I, together with most AGNs and LINERs.

Both diagnostic diagrams suggest that the [O~IV] emission in Holmberg~II is dominated by the ULX 
and that the surrounding star forming regions provide little contamination to this line. 
While this comparison provides good evidence that the line ratios for the ULX are very similar to AGN, 
we note that AGN are different than ULXs, both in their intrinsic emission and their environment.  
For example, AGN emit copiously in the UV and optical, and are surrounded by gas at high densities.

\citet{das08} found a good correlation between the luminosity of the [O~IV] line and the mass of the central BH in standard AGN. 
We estimate a BH mass of 3.5$\times$10$^{4}$~M$_{\odot}$ if this correlation is extrapolated to the Holmberg~II ULX. 
We can also estimate the BH mass using the inner disk temperature from the Modified PLMCD model (0.38~keV).  
We obtain a BH mass of 994~M$_\odot$. However, the flux of the MCD disk component does not dominate the X-ray emission 
(is only 49\% of the total flux). This scenario would imply that the ULX is not in a typical high state,
and therefore the inner disk temperature cannot be used reliably to estimate the mass of the BH \citep{wil06}. 
Finally, using the bolometric luminosity calculated from our SED (1.34~$\times$~10$^{40}$ erg~s$^{-1}$),
we estimate that the lower limit to the mass of the central BH, within the Eddington regime, is 85~M$_\odot$.

\section{CONCLUSIONS}

The detection of the [O~IV] line from the Holmberg~II ULX indicates strong UV and soft X-ray
emission. This line is commonly used to quantify the AGN power in Seyfert galaxies. 
The ratios of this line to lower-ionization lines are comparable to those in AGNs 
and suggest that the [OIV] emission in Holmberg~II results from the ULX 
and not from the surrounding star forming regions. 
The mid-IR line ratios indicate that the intrinsic luminosity of the ULX must be very high. 

The SED constructed based on the X-ray spectral modeling, UV and infrared photometry, 
and previously published radio and optical data suggest that the ULX companion is likely similar to a B2Ib supergiant.
The accretion disk is not expected to contribute very much in the optical.
The overall shape of the SED seems to favor the accretion disk emission as the dominant 
component in the soft X-ray and far UV, and is therefore likely responsible for the photoionization.
In the accompanying paper (Paper~2) we perform detailed photoionization modeling 
to provide further evidence that the detected [O~IV] line is consistent with the ionizing emission from the ULX,
and to better constrain the three X-ray models presented here.

\acknowledgments

C. T. B. is grateful to Lisa Winter for allowing us to use the processed {\it XMM-Newton} data.
He thanks Richard Mushotzky, Lisa Winter and Marcio Mel{\'e}ndez for helpful discussions, 
R. P. D. gratefully acknowledges financial support from the NASA Graduate Student Research Program.
This research has made use of the NASA/IPAC Extragalactic Database (NED) 
which is operated by the Jet Propulsion Laboratory, California Institute of Technology, 
under contract with the National Aeronautics and Space Administration.
This work is based on observations made with the Spitzer Space Telescope, 
which is operated by the Jet Propulsion Laboratory, California Institute of Technology under a contract with NASA.
SMART was developed by the IRS Team at Cornell University and is available through the Spitzer Science Center at Caltech. 
We thank the referee for very helpful and constructive comments that have significantly improved this paper.

\clearpage


\begin{deluxetable}{c|ccc|ccc}
\tablecolumns{6}  
\tablewidth{0pt} 
\tabletypesize{\tiny}
\setlength{\tabcolsep}{0.05in}
\tablenum{1} 
\tablecaption{Measured infrared lines\label{table1}}   
\tablehead{
\colhead{} & \multicolumn{3}{c}{Standard Aperture} & \multicolumn{3}{c}{Small Aperture} \\
\tableline
\colhead{Line} & \colhead{Flux} & \colhead{S/N Ratio} & \colhead{L} & \colhead{Flux} & \colhead{S/N Ratio} & \colhead{L} \\
\colhead{(1)} & \colhead{(2)} & \colhead{(3)} & \colhead{(4)} & \colhead{(5)} & \colhead{(6)} & \colhead{(7)} \\
}
\startdata

$[$Ne II$]$  12.81~$\mu$m  &  $<$6.08	     &  ...   &  $<$0.68        &  $<$1.98	  &	...  &	$<$0.22	        \\
$[$Ne III$]$ 15.56~$\mu$m  &  12.15$\pm$3.4  &  4.60  &  1.35$\pm$0.39  &  5.46$\pm$1.37  &	7.6  &	0.61$\pm$0.15	\\
$[$S III$]$  18.71~$\mu$m  &  2.64$\pm$1.14  &  4.64  &  0.29$\pm$0.13  &  3.02$\pm$1.16  &	4.7  &	0.34$\pm$0.13	\\
$[$Ne V$]$   24.32~$\mu$m  &  $<$3.88	     &  ...   &  $<$0.43        &  $<$1.16	  &	...  &	$<$0.13	        \\
$[$O IV$]$   25.89~$\mu$m  &  7.01$\pm$1.60  &  10.16 &  0.78$\pm$0.2   &  3.63$\pm$0.91  &	44.4 &	0.40$\pm$0.1	\\
$[$S III$]$  33.48~$\mu$m  &  $<$4.39	     &  ...   &  $<$0.49        &  $<$1.53	  &	...  &	$<$0.17	        \\
$[$Si II$]$  34.82~$\mu$m  &  21.91$\pm$5.48 &  9.90  &  2.44$\pm$0.61  &  5.87$\pm$1.47  &	7.2  &	0.65$\pm$0.16	\\

\enddata

\tablecomments{   
We show line fluxes for the standard (slightly smaller than the SH map) 
and the small aperture (4 LH pixels) as defined in Section~2. 
Fluxes are in 10$^{-22}$~W~cm$^{-2}$, luminosities (L) in 10$^{37}$~erg~s$^{-1}$.
If the measurements errors are smaller than the absolute calibration accuracy of 25\%, 
the latter were used. For nondetections we show 3$\sigma$ upper limits.
}
\end{deluxetable}


\begin{deluxetable}{l|cccccc}
\tablecolumns{6}  
\tablewidth{0pt} 
\tabletypesize{\tiny}
\setlength{\tabcolsep}{0.05in}
\tablenum{2} 
\tablecaption{X-ray model fits\label{table2}}   
\tablehead{
\colhead{Model} & \colhead{N$_H$} & \colhead{kT$_{in}$/$\Gamma$$_1$} & \colhead{$\Gamma$/$\Gamma$$_2$} & \colhead{$\Delta\chi^2$/dof} & \colhead{L} & \colhead{MCD flux} \\
\colhead{(1)} & \colhead{(2)} & \colhead{(3)} & \colhead{(4)} & \colhead{(5)} & \colhead{(6)} & \colhead{(7)} \\
\colhead{} & \colhead{(10$^{20}$ cm$^{-3}$)} & \colhead{(keV)} & \colhead{} & \colhead{} & \colhead{(10$^{40}$ erg s$^{-1}$)} \\
}
\startdata

Modified PLMCD	   &  3.26$\pm$0.56  &	0.38$\pm$0.02  &  2.54$\pm$0.02	 &  1.02/135, 0.974/713	&  1.11 &  0.49 \\
PLMCD	           &  10.4$\pm$0.70  &	0.27$\pm$0.02  &  2.42$\pm$0.05	 &  1.1386/849	        &  2.28 &  0.12 \\
BPL                &  5.49$\pm$0.78  &	1.73$\pm$0.10  &  2.57$\pm$0.02	 &  1.0316/849	        &  1.41 &  0.60 \\

\enddata

\tablecomments{  
(1): X-ray model, as described in Section~2.2.
(2): Intrinsic hydrogen column density. 
The Galactic column density from KWZ (3.42$\times$10$^{20}$ cm$^{-2}$) was added separately. 
(3): Inner disk temperature for the MCD component. For the broken power-law model (BPL), this is 
the first (low energies) photon index parameter. The break is at 1.0~keV.
(4): Photon index for the power-law component. 
For the broken power-law model this is the second (high energies) photon index parameter.
(5): Reduced $\chi^2$ values for the fit and the number of degrees of freedom. 
For the Modified PLMCD model, we quote two values for each separately fitted component (see text for details).
(6): Unabsorbed (intrinsic) luminosities between 0.1 and 10 keV.
(7): MCD component unabsorbed flux as fraction of the total flux for the PLMCD model.
For the other two models, this simply the ratio of the flux between 0.1 and 1.0~keV
to the flux in the whole range (0.1$-$10~keV)
}
\end{deluxetable}


\begin{deluxetable}{ccccccc}
\tablecolumns{6}  
\tablewidth{0pt} 
\tabletypesize{\tiny}
\setlength{\tabcolsep}{0.05in}
\tablenum{3} 
\tablecaption{Multi-wavelength photometry measurements\label{table3}}   
\tablehead{
\colhead{Instrument} & \colhead{Program (observation)} & \colhead{Wavelength} & \colhead{Date} & \colhead{Flux Density} & \colhead{A$_{\lambda}$} & \colhead{Corrected Fux} \\
\colhead{(1)} & \colhead{(2)} & \colhead{(3)} & \colhead{(4)} & \colhead{(5)} & \colhead{(6)} & \colhead{(7)}\\
}
\startdata
\multicolumn{7}{c}{Photometry performed in this paper}\\  
\tableline

{\it Spitzer} IRAC     & SINGS 159  &   3.6~$\mu$m   & 2004 Oct 10 & 0.15$\pm$0.03 mJy     &  ...   &   ... \\
{\it Spitzer} IRAC     & SINGS 159  &   8.0~$\mu$m   & 2004 Oct 10 & 0.26$\pm$0.05 mJy     &  ...   &   ... \\
{\it Spitzer} MIPS     & SINGS 159  &   24.0~$\mu$m  & 2004 Oct 14 & 0.69$\pm$0.14 mJy     &  ...   &   ... \\
{\it XMM-Newton} OM    & 0200470101 &   UVW1 2910\AA & 2004 Apr 15 & 33.1$\pm$0.1 $\mu$Jy  &  0.60  &  57.5 $\mu$Jy \\
{\it GALEX} NUV        & NGS        & 1750$-$3013\AA & 2003 Dec 5  & 26.3$\pm$0.2 $\mu$Jy  &  0.95  &  63.1 $\mu$Jy \\
{\it GALEX} FUV        & NGS        & 1350$-$2013\AA & 2003 Dec 5  & 33.7$\pm$0.3 $\mu$Jy  &  0.84  &  73.0 $\mu$Jy \\

\tableline                                                                                                 
\multicolumn{7}{c}{Previously published photometry}\\  
\tableline

VLA                    & AT159      & 1.4 GHz        & 1994 Apr 3  & 1.174$\pm$0.085 mJy   &  ...   &   ... \\
VLA                    & AR165, AR227, AR258  & 4.86 GHz  & 1990 Jul 21 $-$ 1991 Dec 6 & 0.667$\pm$0.207 mJy  & ...   &   ... \\
HST ACS                & GO 9684    & F550M          & 2002 Nov 24 & 6.56$\pm$0.03 $\mu$Jy & 0.32   &  9.51 $\mu$Jy \\

\enddata

\tablecomments{   
All the photometric measurements performed in this paper used an aperture of 50~pc diameter (3.3$\arcsec$).
These values represent the integrated fluxes over this region, including the nebular emission
and are contaminated by other nearby sources, therefore we consider them as upper bounds.
The published data is taken from \citet{mill05} (radio {\it VLA}) and from KVZ (optical {\it HST}).
The radio emission is extended, and is comparable to our extraction region. 
The optical measurement is for the star-like source which is believed to be the companion of the accreting BH. 
This measurement is likely less contaminated by the nebular emission or other sources.
The extinctions in column 6 are both Galactic and local from KWZ.
For the infrared an radio data the extinctions are negligible.
}
\end{deluxetable}


\begin{figure}
\epsscale{0.9}
\plotone{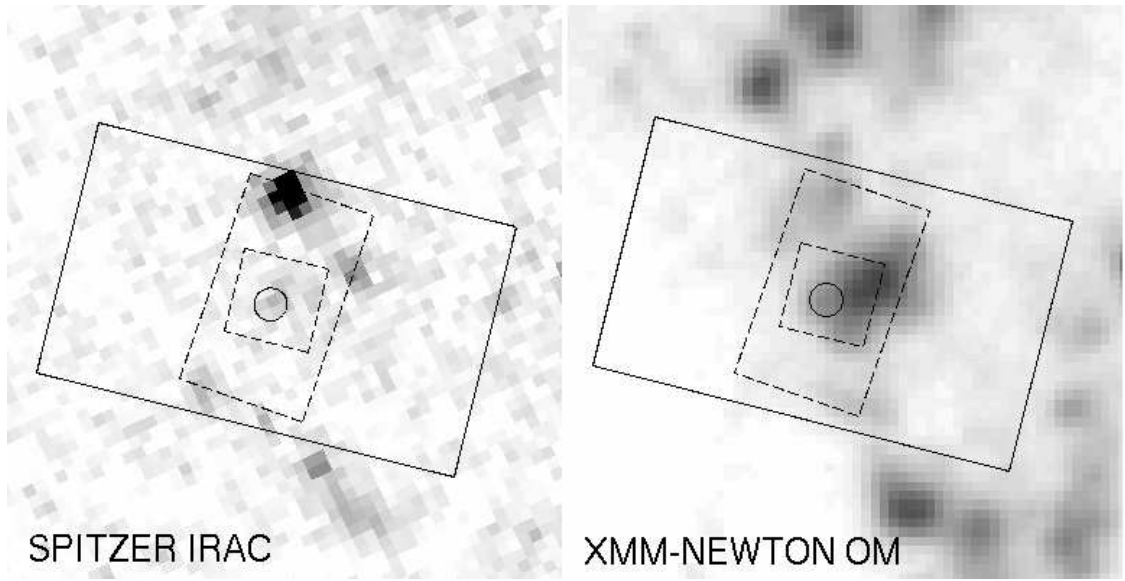}
\caption{
{\it Spitzer} IRAC (8~$\mu$m filter, left) and  {\it XMM-Newton} OM (UVW1 filter, right) images 
of the Foot nebula. The aperture used for the ULX photometry is shown as a small circle.
At the Holmberg~II distance, it has a diameter of 50~pc.
With dotted lines we show the two apertures used to extract the IRS spectra shown in Fig.~\ref{spectra}.
The larger aperture (continuous line) corresponds to the whole LH map.
Notice that the bright star which contaminates the infrared emission in Fig.~\ref{maps} 
is very faint in the UV image. The strongest UV emission in the nebula 
is located west of the ULX and it originates from the OB association better seen in Fig.~\ref{maps} .
}
\end{figure}

\begin{figure}
\epsscale{0.6}
\plotone{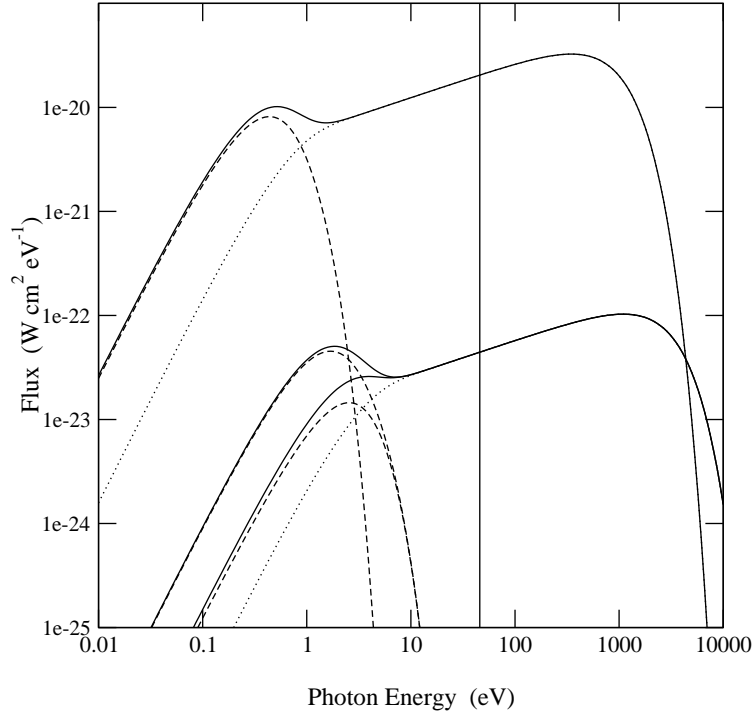}
\caption{
Examples of irradiated disks spectra. The higher flux spectrum is calculated for a 
1000~M$_{\odot}$ intermediate-mass BH binary and an outer radius of 10$^{14}$~cm.
The lower spectrum is for a stellar-mass BH binary of 10~M$_{\odot}$ and two disk sizes:
3$\times$10$^{11}$~cm and 10$^{12}$~cm. The continuous lines show the total spectrum,
dotted lines the spectrum from inner disk, not affected by the irradiation, 
and the dashed lines from the outer part. The vertical line marks the [O~IV] ionization edge.
The reprocessed photons have energies below 10~eV, and therefore not ionizing.
All cases assume accretion at Eddington luminosity. 
}
\end{figure}

\begin{figure}
\epsscale{0.7}
\plotone{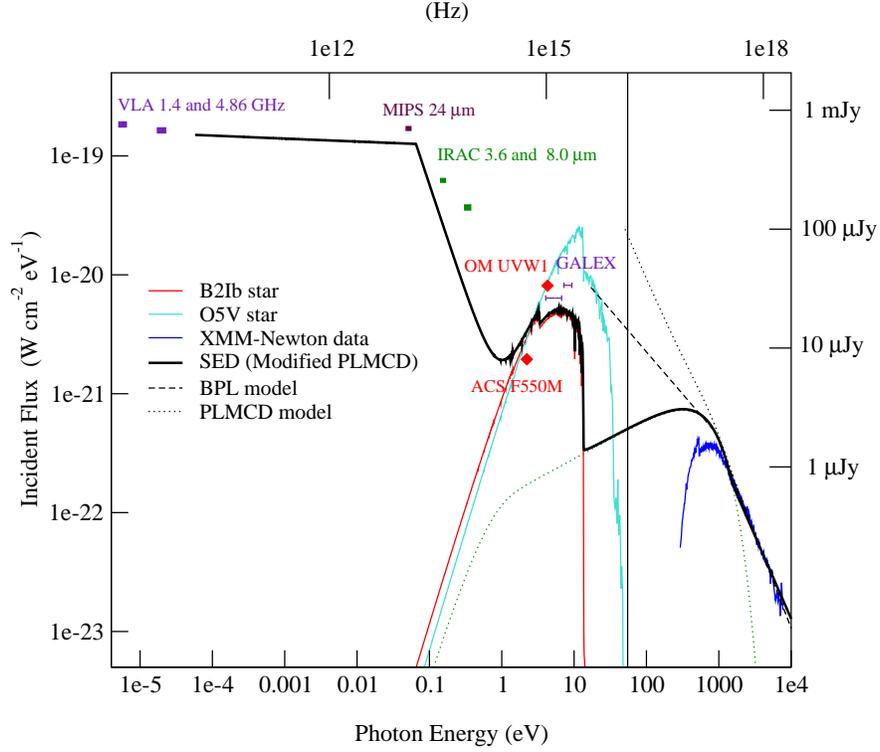}
\caption{
This figure shows the SED constructed in Section~4.
The radio data is from \citet{mill05}, the V-band magnitude for the optical counterpart of the ULX is quoted from KWZ.
Our measured fluxes from IRAC, MIPS, OM and {\it GALEX} are upper bounds, and therefore our model 
is below these data points.
The two stellar spectra correspond to the spectral types range consistent with the colors and magnitudes in KWZ. 
The X-ray data is from  the long ($\sim$100~ks) {\it XMM-Newton} observation ID 0200470101.
The Modified PLMCD model was obtained by fitting the MCD and power-law components separately as explained in the text.
The other two X-ray models, PLMCD and BPL are also shown extrapolated to lower energies, 
they clrealy overpredict the UV and optical fluxes.
The [O~IV] edge at 54.93~eV is shown as a vertical line.
}
\end{figure}

\begin{figure}
\epsscale{1}
\plottwo{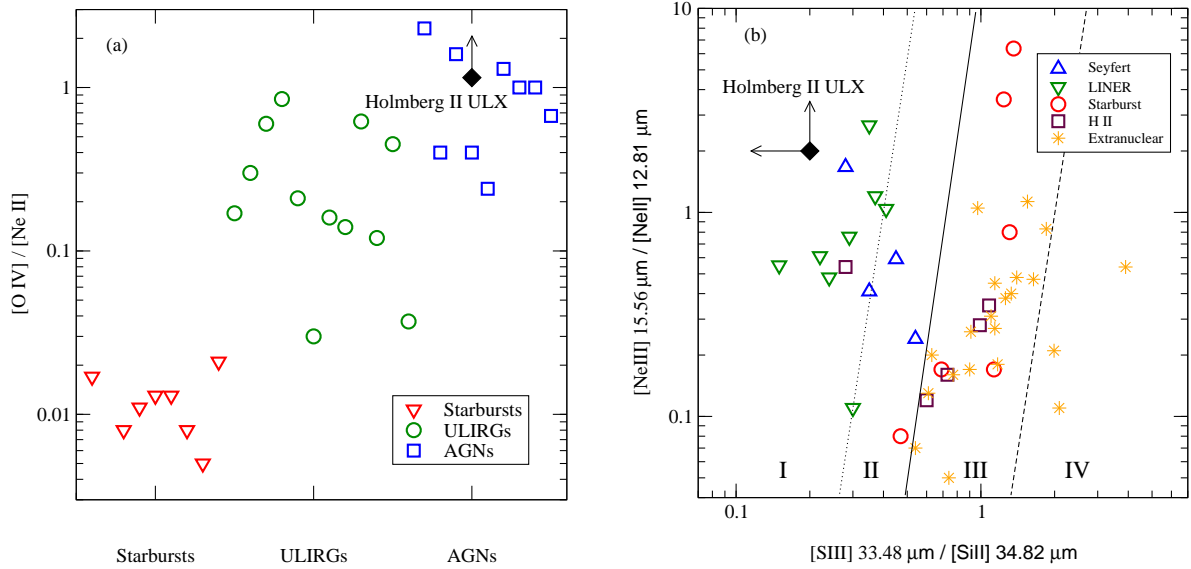}{figure7b.eps}
\caption{
Ratio line diagnostics comparison between Holmberg~II ULX and other extragalactic ionizing sources: AGN, starburts, etc.
Its position in both plots indicate a source very similar to AGNs and LINERs.
a) Diagnostic that uses the [O~IV] line normalized to the [Ne~II] line.
The AGN, starburts and ULIRG data is taken from \citet{gen98}. 
Many of the starburts and ULIRGs are actually upper limits \citep[see Fig.~3 of][]{gen98}.
For Holmberg~II ULX we show the lower limit because [Ne~II] was not detected.
b) Diagnostic plot that uses the [S~III] and [Si~II] lines to distinguish between different sources.  
We plot all the data presented by \citet{dale06} (their Fig.~5), which includes: Seyferts, LINERs, starburt galaxies, 
H~II regions and extranuclear star-forming regions.
For Holmberg~II ULX we show the lower and upper limits because [Ne~II] and [S~III], respectively, were not detected.
} \label{ratios}
\end{figure}

\end{document}